# Optimal Selection of Assets Investing Composition Plan based on Grey Multi Objective Programming method

*Gol Kim[a], Ri Suk Yun[b]*

[a] Center of Natural Sciences, University of Sciences, Kwahakdong-1, Unjong District, Pyongyang, DPR Korea (E-mail:golkim124@yahoo.com)

[b] Foreign Economic General Bureau, Pyongyang, DPR Korea

**Abstract-** The problem for selection of appropriate assets investing composition projects such as assets rationalization plays an important role in promotion of business systems.

We consider the assets investing composition plan problems subject to grey multi objective programming with the grey inequality constraints.

In this paper, we show in detail the entire process of the application from modeling the case problem to generating its solution.

To solve the grey multi objective programming problem, we then develop and apply an algorithm of grey multiple objective programming by weighting method and an algorithm of grey multiple objective programming based on $\theta$-positioned programming method.

These algorithms all regard as of great importance uncertainty (greyness) at grey multi objective programming and simple and easy the calculating process. The calculating examples of paper also show ability and effectiveness of algorithms.

## 1. Introduction

Recently, there are many papers for the investment project selection modeling (for instances; Mikulas Luptacik (2012), Song Ye-Xing, Chen Mian-Yun (2002), Elton.E.J. and Gruber, M.J (1995)).
Safety first portfolio choice based on financial and sustainability returns has been researched by Gregor Dorfleitner, Sebastian Utz (2012) and the problem of selecting the most economic project under uncertainty using bootstrap technique and fuzzy simulation has been studied by Kamran Shahanaghi et al. (2012) .Interactive multi objective optimization approach to the input–output design of opening new branches has been suggested by K. Sam Park, Dong Eun Shin (2012).
Research of transportation investment project selection using fuzzy multi objective programming has been progressed by Junn-Yuan Teng, et al. (1998).
Preference elicitation from inconsistent judgments using multi-objective optimization has been studied by S. Siraj, L. Mikhailov, J.A. Keane (2012) and a conceptual methodology for transportation projects selection has been proposed by Iman Nosoohi, et al. (2011).
Robust optimization approach to R&D project selection, Mohammad Modarres has been researched by Farhad Hassanzadeh (2009). Mixed-integer linear programming for resource leveling problems has been studied by Julia Rieck, et al (2012), and Multi-objective evolutionary algorithm for donor recipient decision system in liver transplants has been researched by Manuel Cruz-Ramírez , et al. (2012). But, these results are the fuzzy programming methods.
In this paper, we consider the decision-making for assets investing composition plan by method of grey multi object decision making.
Let us consider the problem which investors determine a selected plan to the market with some assets, that is, a stock, a public debt, credit, and so on.

We assume that selection is offered to market with some kind of assets $S_i \ (i=1,2,\cdots,n)$ by investor. Suppose that total amount of money $M$ possessed by a company is considerably large and it can be used to invest in some period.

These $n$ kind of asserts $S_i$ was evaluated by financial analyst of this company

So, they take out calculate that average profit rate to purchase the asset $S_i$ is given by grey number $\otimes r_i$. And they take out predict that damage rate by risk is given by grey number $\otimes q_i$ in this period.

Considering that invest the more dispersed, total risk the less, the company have confirmed that total risk is a quantity causing the greatest danger of $S_i$ for investing plans.

When purchasing the asset $S_i$, it is demand to disburse transaction cost and consumption rate given by grey number $\otimes p_i$. The purchase of asset $S_i$ is not exceeding a given grey number $\otimes u_i$.

Besides, if interest rate by bank savings is given by grey number $\otimes r_0$ in the same period, then they are non- transaction and no-risk.

In that case, let us consider the company's decision- making plans of invest project composition. Our purpose is that make the total risk as possible smaller while make the profit as possible greater by the purchase of asserts or bank savings interest.

The factor affecting the decision making plan of invest project is many-sided.

Accordingly economic information data obtained through the collection are incomplete and uncertain. Therefore, the system of the decision making for investing composition plan is a grey system.

The essential question of investing project is to achieve the composition in which the profit and the safety are much satisfactory.

## 2. Mathematical description for assets investing composition plan

In grey theory, random variables are regarded as grey numbers, and a stochastic process is referred to as a grey process. A grey system is defined as a system constraining information presented as grey numbers, and a grey decision is defined as a decision made within a grey system (Liu Si-feng, Lin Yi (2004)).

**Definition 1.** Let $x$ denote a closed and bounded set of real numbers. A grey number $\otimes x$ is defined as an interval with known upper and lower bounds but unknown distribution information for $x$

$$\otimes x = [\underline{\otimes}x, \overline{\otimes}x] = \{x' \in x \mid \underline{\otimes} \leq x' \leq \overline{\otimes}\}$$

where $\underline{\otimes}x$ are the lower and $\overline{\otimes}x$ upper bounds of $\otimes x$, respectively

**Definition 2.** The whitened value $\tilde{\otimes}x$ of a grey number $\otimes x$ is defined as a deterministic value with its value lying between the lower and upper bounds of $\otimes x$. It can be marked by:
$\tilde{\otimes}x = x', \ x' \in [\underline{\otimes}x, \overline{\otimes}x]$

We assume that bank savings is also a sort of investment item.

Let $S_0$ denote the bank savings and $\otimes q_0$ and $\otimes p_0$ denote risk expense rate and transaction cost rate corresponding to $S_0$, respectively. Then $\otimes q_0 = 0, \otimes p_0 = 0$.

If addition transaction cost invested in i'th sorts assert is $A_i \ (i=0,1,2,\cdots,n)$, then we have such as

$$A_i = \otimes p_i \times \max(Mx_i, u_i f(x_i))$$

where

$$f(x_i) = \begin{cases} 0 : x_i \neq 0 \\ 1 : x_i > 0 \end{cases}$$

Above express means that if it is not invested in assert $S_i$ then $x_i = 0$ and if it is invested in assert $S_i$ then we take the large among $Mx_i$ and $u_i$.

Thus, the total amount of investment M is denoted as following;

$$M = \sum_{i=0}^{n} A_i + \sum_{i=0}^{n} Mx_i$$

where $\sum_{i=0}^{n} A_i$ is the total amount of additional transaction cost $\sum_{i=0}^{n} Mx_i$ is the total amount of finance consumed to purchase each asserts.

On the one hand pure profit is denoted as following;

$$Z_1 = \sum_{i=0}^{n} \otimes r_i M x_i - \sum_{i=0}^{n} A_i = \sum_{i=0}^{n} (\otimes r_i + 1) M x_i - M$$

The total amount of risk is denoted such as;

$$Z_2 = \max_{0 \leq i \leq n}(\otimes q_i x_i M)$$

Seeing that, we obtain programming model with two objects;

$$\max Z_1 = \sum_{i=0}^{n} (\otimes r_i + 1) M x_i - M \qquad (2.1)$$

$$\min Z_2 = \max_{0 \leq i \leq n}(\otimes q_i x_i M) \qquad (2.2)$$

$$\text{Subject:} \begin{cases} \sum_{i=0}^{n} A_i + \sum Mx_i = M \\ x_i \geq 0 \end{cases} \qquad (2.3)$$

where

$$A_i = \otimes p_i \times \max(Mx_i, u_i f(x_i)) \qquad (2.4)$$

$$f(x_i) = \begin{cases} 0 : x_i \neq 0 \\ 1 : x_i > 0 \end{cases}$$

The problem determining assets investing plan expressed by above (2.1)-(2.4) is a two-objective nonlinear programming problem. It can be denote in general form such as:

$$Min\{-Z_1(x), Z_2(x)\}, x \in X$$

where $x$ is the set of decision variables; X is the set of feasible points defined by given constraints, i.e. Eqs. (2.3), (2.4).

Eqs (2.1) and (2.2) are the two objective functions, respectively, to be minimized.

Directly applying the notion of optimality for single-objective nonlinear programming to this two-objective nonlinear programming allows us to arrive a complete optimal solution that simultaneously minimizes these two objective functions. However, in general, such a complete optimal solution does not always exist when the objective functions conflict with each other (Sakawa, M (1993)).

In our problem, these two objectives conflict with each other. Consequently, instead of complete optimal solution, the Pareto optimality is the solution where no objective can be reached without simultaneously worsening at least one of the remaining objectives.

The Pareto optimal solutions can be solved by the constraint method for our two-objective programming. The constraint method for characterizing Pareto optimal solutions attempt method to solve the following constraint problem formulated by taking one objective function $Z_1(x)$ as the objective function and allowing the other objective function, $Z_2(x)$, to be an inequality constraint for some selected values of $e_2$ ( Sakawa, M(1993), Steuer, P.E(1986),)

$$Min \quad \{-Z_1(x)\}$$
$$s.t. \begin{cases} Z_2(x) \leq e_2 \\ x \in X \end{cases} \qquad (5)$$

The relationships between the optimal solution $x^*$ to the constraint problem and the Pareto optimality of the two-objective programming problem have been proven to follow the theorem(Sakawa, M(1993)),such that $x^* \in X$ is a Pareto optimal solution of the two –objective nonlinear programming problem, if and only if $x^*$ is an optimal solution of the constraint problem for some $e_2$.

Consider the Lagrange function, $L(x,\lambda) = Z_1(x) + \lambda_{12}(Z_2(x) - e_2)$ for the constraint problem with respect to the e-constraints. If the Lagrange multiplier, $\lambda_{12}$ associated with the active constraint, i. e. $Z_2(x) - e_2 = 0$, then the corresponding Lagrange multiplier can be proven to lead to the trade off rates between $Z_1(x)$ and $Z_2(x)$ by $\lambda_{12} = -\partial Z_1(x)/\partial Z_2(x)$. The trade-off rate means the marginal decrease of $Z_1(x)$ with one unit increase in $Z_2(x)$. Herein we use $\lambda_{12} = -\Delta Z_1(x)/\Delta Z_2(x)$ to approximate $-\partial Z_1(x)/\partial Z_2(x)$.

Then, both Pareto optimal solutions and trade-off rates can be obtained by altering the values of $e_2$ and solving the corresponding constraint problems .In this manner, a variety of frequency plans for routes can be generated from Pareto optimal solution for decision makers. These Pareto optimal solutions can be plotted as a Pareto optimal boundary. Along this Pareto optimal boundary, this study attempts to obtain a solution nearest to the ideal point. The ideal point is defined as the point $Z^{ideal} - (Z_2^{\min}, Z_1^{\min})$, where $Z_1^{\min}$ and $Z_2^{\min}$ are the values of the objective function for single-objective programming that minimize $Z_1(x)$ and $Z_2(x)$, respectively. Realizing the ideal point is generally infeasible; Zeleny ([11]) introduced the concept of compromise programming. The compromise solution is a Pareto optimal solution which has the shortest geometrical distance from the ideal point. In the following case study, compromise programming is applied to determine and derive a compromise solution from these Pareto optimal solutions.

But, validity of such results are not obvious in the case with nonlinear object (2) .Besides the coefficients of given problem are the grey numbers. Therefore, to solve of this problem is very annoying. C.W.Duin, A.Volgenant (2012) have been introduced the on weighting two criteria with a parameter in combinatorial optimization problems. But this method don't too appropriate for our problem.

So, we take another method to convert for nonlinear object (2.2) into linear object. One of these methods is a method with adding linear weight.

Now, if $\otimes\lambda$ is total amount of invest risk, then $1 - \otimes\lambda$ is total amount of pure profit, where $\otimes\lambda \in [0,1]$. Grey number $\otimes\lambda$ is given by decision making of investor.

Then single-objective nonlinear programming is given by the linear programming with adding linear weight such as:

$$Min \, Z = (1 - \otimes\lambda) \cdot (-Z_1) + \otimes\lambda \cdot Z_2 \qquad (2.6)$$

Now, to obtain the convenience form of risk function we put $x_{n+1} = Z_2$. Then, we obtain $q_i x_i \leq x_{n+1}, (i=0,1,\cdots,n)$

The optimal solution has got to at place which $\max_{1 \leq i \leq n}(\otimes q_i x_i)$ arrival in $x_{n+1}$. Thus finally, we have single-objective grey linear programming:

$$\min Z = (1 - \otimes\lambda) \times [-(\sum_{i=0}^{n}(\otimes r_i + 1)x_i - 1)] + \otimes\lambda \cdot x_{n+1} \qquad (2.7)$$

$$\text{s.t.} \begin{cases} \sum_{i=0}^{n}(1+\otimes p_i) = 1 \\ x_i \geq 0 \\ \otimes q_i x_i - x_{n+1} \leq 0 \end{cases} \quad (2.8)$$

We can reformulate the grey linear programming (7), (8) in to standard form as following;

$$Min\ Z = C(\otimes)X \quad (2.9)$$

$$s.t. \begin{cases} A(\otimes)X \leq b(\otimes) \\ X \geq 0 \end{cases} \quad (2.10)$$

where

$$C(\otimes) = [c_1(\otimes), c_2(\otimes), \cdots, c_{n+1}(\otimes)], \ b(\otimes) = [b_1(\otimes), b_2(\otimes), \cdots, b_m(\otimes)]^T$$

$$A(\otimes) = \begin{bmatrix} a_{11}(\otimes), a_{12}(\otimes), \cdots, a_{1,n+1}(\otimes) \\ a_{21}(\otimes), a_{22}(\otimes), \cdots, a_{2,n+2}(\otimes) \\ \cdots \quad \cdots \quad \cdots \\ a_{m1}(\otimes), a_{m2}(\otimes), \cdots, a_{m,n+1}(\otimes) \end{bmatrix}$$

here, $c_j(\otimes) \in [\underline{c_j}, \overline{c_j}]$, $\underline{c_j}(\otimes) \geq 0$, $b_j(\otimes) \in [\underline{b_j}, \overline{b_j}]$, $\underline{b_j} \geq 0$, $a_{ij}(\otimes) \in [\underline{a_{ij}}, \overline{a_{ij}}]$, $\underline{a_{ij}}(\otimes) \geq 0$
$(i = 1, 2, \cdots, m; j = 0, 1, 2, \cdots, n+1)$

Then, we defined that the problem (2.9), (2.10) is grey parameter linear programming and it denote LPGP.

And, we defined that $C(\otimes)$ is grey price vector, $A(\otimes)$ is grey consumption matrix, $b(\otimes)$ is grey resources constraint vector, X is decision making vector, respectively .X is also grey Vector.

## 3. Grey Drifting type linear programming

**Definition 3.** Let $\rho_j, \beta_i, \delta_{ij} \in [0,1]$ $(i = 1, 2, \cdots, m; j = 0, 1, 2, \cdots, n+1)$. Let

$$\tilde{c}_j(\otimes) = \rho_j \overline{c_j} + (1-\rho_j)\underline{c_j}$$
$$\tilde{b}_j(\otimes) = \beta_j \overline{b_j} + (1-\beta_j)\underline{b_j}$$
$$\tilde{a}_{ij}(\otimes) = \delta_{ij} \overline{a_{ij}} + (1-\delta_{ij})\underline{a_{ij}}$$

are the white values of the grey parameters, respectively. We denote that $\tilde{C}(\otimes)$ is white price vector, $\tilde{A}(\otimes)$ is white consumption matrix, and $\tilde{b}(\otimes)$ is white resources constraint vector, respectively. Then positioned programming of LPGP is defined as;

$$Min\ Z = \tilde{C}(\otimes)X \quad (3.1)$$

$$s.t. \begin{cases} \tilde{A}(\otimes)X \leq \tilde{b}(\otimes) \\ X \geq 0 \end{cases} \quad (3.2)$$

We call that $\rho_j$ is positioned coefficient of price, $\beta_i$ is positioned coefficient of resource restrict $\delta_{ij}$ is positioned coefficient of consumption, respectively.

**Proposition.** The optimal value max Z of the positioned programming of LPGP is concerned with m+(n+1)+m(n+1) dimension variables $\rho_j, \beta_i, \delta_{ij}$. Therefore we can descried the optimal value max Z of the positioned programming of LPGP as function of variables of $\rho_j, \beta_i, \delta_{ij}$;

$$Min\ Z = Z(\rho_j, \beta_i, \delta_{ij}) \qquad (3.3)$$

Similarly, we can denote the positioned programming of LPGP as following;
$LP((\rho_j, \beta_i, \delta_{ij}))$.

**Definition 4.** For arbitrary $i = 1, 2, \cdots, m; j = 0, 1, 2, \cdots, n+1$, if we have; $\rho_j = \rho, \beta_i = \beta, \delta_{ij} = \delta$, then the positioned programming corresponding to it is called $(\rho, \beta, \delta)$ positioned programming and it is denoted as $LP(\rho, \beta, \delta)$. The optimal value of $LP(\rho, \beta, \delta)$ is denoted by $\min Z(\rho, \beta, \delta)$

Theorem1. The positioned programming of LPGP has the properties such as:

1°. If $\rho = \rho_0, \beta = \beta_0, \delta_1 \leq \delta_2$, then $\max Z(\rho_0, \beta_0, \delta_1) \geq \max Z(\rho_0, \beta_0, \delta_2)$

2°. If $\rho_1 \leq \rho_2, \beta = \beta_0, \delta = \delta_0$, then $\max Z(\rho_1, \beta_0, \delta_0) \leq \max Z(\rho_2, \beta_0, \delta_0)$

3°. If $\rho = \rho_0, \beta_1 \leq \beta_2, \delta = \delta_0$, then $\max Z(\rho_0, \beta_1, \delta_1) \leq \max Z(\rho_0, \beta_2, \delta_2)$

**Definition 5.** Let $\rho = \beta = 1, \delta = 0$. Then, the positioned programming LP (1, 1, 0) corresponding to it is called ideal model of LPGP and optimal value of LP (1, 1, 0) is denoted by $\max \overline{Z}$.

Let $\rho = \beta = 0, \delta = 1$. Then, the positioned programming LP (0, 0, 1) corresponding to it is called critical model of LPGP and optimal value of LP (0, 0, 1) is denoted by $\max \underline{Z}$.

Let $\rho = \beta = \delta = \theta$. Then, the positioned programming LP ($\theta, \theta, \theta$) corresponding to it is called $\theta$-positioned programming model of LPGP and optimal value of LP ($\theta$) is denoted by $\max Z(\theta)$.

Let $\theta = 0.5$. Then, the $\theta$-positioned programming LP (0.5) corresponding to it is called mean white programming model of LPGP

For $\rho, \beta, \delta, \theta \in [0,1]$, the positioned programming $LP(\rho, \beta, \delta)$ of LPGP have the properties such as:

1°. $\max \underline{Z} \leq \max Z(\rho, \beta, \delta) \leq \max \overline{Z}$

2°. $\max \underline{Z} \leq \max Z(\theta) \leq \max \overline{Z}$

**Definition 6.** For given $\rho, \beta, \delta \in [0,1]$ the pleased degree of $LP(\rho, \beta, \delta)$ is defined as;

$$\mu(\rho, \beta, \delta) = \frac{1}{2}(1 - \frac{\max \underline{Z}}{\max Z(\rho, \beta, \delta)}) + \frac{1}{2}\frac{\max Z(\rho, \beta, \delta)}{\max \overline{Z}} \qquad (3.4)$$

The pleased degree of positioned programming of $LP(\rho, \beta, \delta)$ express the relation of the optimal value $\max \underline{Z}$ of critical model and the optimal value $\max \overline{Z}$ of ideal model for optimal value $\max Z(\rho, \beta, \delta)$ of positioned programming. The more approach for $\max Z(\rho, \beta, \delta)$ to $\max \overline{Z}$, more $\mu(\rho, \beta, \delta)$ is great.

Conversely, the more approach for $\max Z(\rho, \beta, \delta)$ to $\max \underline{Z}$, more $\mu(\rho, \beta, \delta)$ is small.

Similarity, we can define the pleased degree of general positioned programming $LP(\rho_j, \beta_i, \delta_{ij})$ and $\theta$-positioned programming $\mu(\theta)$

Proposition. For any $\rho, \beta, \delta \in [0,1]$, we have;

$$0 \leq \mu(\rho, \beta, \delta) \leq 1$$

**Definition 7.** For given $D = [\mu_0, 1]$, if $\mu(\rho, \beta, \delta) \in D$, then the optimal solution of positioned programming $LP(\rho, \beta, \delta)$ is called by pleased solution of LPGP.

## 4. Solving algorithm of general grey multiple objective programming

Generally, the problem of investing composition plan become the grey multiple objective programming.
Grey multiple objective programming belong cross research field of multiple objective programming and grey system theory.

**Definition 8.** We assume that $x = (x_1, x_2, \cdots, x_n)$ is decision making vector and $\otimes_1^{(i)}, \otimes_2^{(j)}, \otimes_3^{(k)}$ $(i = 1, 2, \cdots, m; j = 1, 2, \cdots, p; k = 1, 2, \cdots, q)$ are grey parameters.
Then, the problem

$$\max(\min)(f_1(x, \otimes_1^{(1)}), f_2(x, \otimes_1^{(2)}), \cdots, f_m(x, \otimes_1^{(m)}))$$

$$s.t. \begin{cases} g_j(x, \otimes_2^{(j)}) \leq (\geq) 0, j = 1, 2, \cdots, p \\ h_k(x, \otimes_3^{(k)}) = 0, k = 1, 2, \cdots, q \\ x \geq 0 \end{cases} \quad (4.1)$$

is called general model of grey multiple objective programming problem. Where, $f_i(x, \otimes_1^{(i)})$ is grey objective function and $g_j(x, \otimes_2^{(j)}), h_k(x, \otimes_2^{(k)})$ are grey constrained functions.

In this paper, we consider the solving method of grey multiple objective programming M-1 in the case of all grey parameters are interval type numbers. That is, in the problem M-1, if we put $\otimes_1^{(i)} = (\otimes_{11}^{(i)}, \otimes_{12}^{(i)}, \cdots, \otimes_{1n}^{(i)})$, $\otimes_2^{(j)} = (\otimes_{21}^{(j)}, \otimes_{22}^{(j)}, \cdots, \otimes_{2n}^{(j)})$, $\otimes_3^{(k)} = (\otimes_{31}^{(k)}, \otimes_{32}^{(k)}, \cdots, \otimes_{3n}^{(k)})$, then $\otimes_{1t}^{(i)} = [\underline{c}_{1t}^{(i)}, \overline{c}_{1t}^{(i)}], \otimes_{2t}^{(j)} = [\underline{a}_{2t}^{(j)}, \overline{a}_{2t}^{(j)}]$, $\otimes_{3t}^{(k)} = [\underline{b}_{3t}^{(k)}, \overline{b}_{3t}^{(k)}]$ all are interval numbers.

We should discuss problem (4.1) in the case of maximum value. In the minimum value also discussion is progressed similarly.

### 4.1. Algorithm of grey multiple objective programming by weighting method

This method is called by Algorithm 1
First, we progress mean whiting handling for grey constrained conditions. So, grey parameters of constrained conditions are replaced by using technical coefficients and resource allocation coefficient.
We define admissible domain of grey multiple objective programming with whitened constrained conditions, that is,

$$U = \{x | x \geq 0, \ g_j(x, \otimes_2^{(j)}) \leq 0, h_k(x, \otimes_2^{(k)}) = 0\}.$$

We take arbitrary admissible solution $x^{(t)} = (x_1^{(t)}, x_2^{(t)}, \cdots, x_n^{(t)}) (t = 1, 2, \cdots, l)$ in admissible domain.
Commonly, we take $l$ which satisfied condition; $m \leq l \leq 2m$.
Taking each sub-objective $f_i$ and calculating the value of defined admissible solution, then we obtain interval grey matrix such as;

$$F(\otimes) = (f_i^{(t)}(\otimes))_{m \times l} = (f_i(x^{(t)}, \otimes_1^{(i)}))_{m \times l} \quad (4.2)$$

Where $f_t^t(\otimes) = [\underline{f}_i^t, \overline{f}_i^t]$ $(i = 1, 2, \cdots, m; t = 1, 2, \cdots, l)$.

To improve comparability, we progress the normalization handling for $f_t^t(\otimes)$ by eliminate the effect of magnitude of given data. In order to aim, we define grey extreme difference transformation method such as;
    for the effective type objective value;

$$r_{it} = \frac{f_i^t - f_i^\nabla}{\overline{f}_i^t - \underline{f}_i^\nabla}, \quad \overline{r}_{it} = \frac{\overline{f}_i^t - f_i^\nabla}{\overline{f}_i^t - \underline{f}_i^\nabla} \quad (4.3)$$

For the cost type objective value;

$$\underline{r}_{it} = \frac{f_i^\nabla - f_i^t}{\overline{f}_i^\Delta - \underline{f}_i^\nabla}, \quad \overline{r}_{it} = \frac{\overline{f}_i^\nabla - f_i^t}{\overline{f}_i^\Delta - \underline{f}_i^\nabla} \quad (4.4)$$

where $\overline{f}_i^\nabla = \max_{1 \leq t \leq l}\{\overline{f}_i^t\}$, $\underline{f}_i^\Delta = \min_{1 \leq t \leq l}\{\underline{f}_i^t\}$

Proceeding the normalization handling for objective value in interval grey matrix (4.1) by grey extreme difference transformation formula (4.3), (4.4), we can obtain interval grey matrix by

$$R(\otimes) = (r_{it}(\otimes))_{m \times n} \quad (4.5)$$

where $0 \leq \underline{r}_{it} \leq r_{it}(\otimes) \leq \overline{r}_{it} \leq 1 (i = 1, 2, \cdots, m; t = 1, 2, \cdots, l)$

Let calculate the deviation of each interval grey number in matrix (19). That is, se denote deviation of admissible solution $x(t)$ and other all admissible solutions under sub-objective by $D_{it}$

$$D_{it} = \sum_{s=1}^{l} d(r_{it}(\otimes), r_{is}(\otimes)), \quad i = 1, 2, \cdots, m; t = 1, 2, \cdots, l$$

$$D_i = \sum_{t=1}^{l} D_{it} = \sum_{t=1}^{l}\sum_{s=1}^{l} d(r_{it}(\otimes), r_{is}(\otimes)), i = 1, 2, \cdots, m; t = 1, 2, \cdots, l \quad (4.6)$$

where $d(r_{it}(\otimes), r_{st}(\otimes)) = |\underline{r}_{it} - \underline{r}_{st}| + |\overline{r}_{it} - \overline{r}_{st}|$.

$D_i$ denotes total deviation of given admissible solution and order admissible under the sub-objective $f_i$.

Calculating output entropy of sub-objective function, then we obtain

$$E_i = -(\ln l)^{-1} \sum_{t=1}^{l} \frac{D_{it}}{D_i} \ln \frac{D_{it}}{D_i}, i = 1, 2, \cdots, m \quad (4.7)$$

Easily, we can know that $0 \leq E_i \leq 1$.

Calculating weight of sub-objective function, we obtain;

$$w_i = \frac{1 - E_i}{\sum_{i=1}^{m}(1 - E_i)}, i = 1, 2, \cdots, m \quad (4.8)$$

Remark. To obtain more accurate weight value of sub-objective, decision maker can offer preference(subjective weight coefficient) of each sub-objective $\mu_i$ or can take another set of admissible solutions in admissible domain .By repeating above-mentioned calculating procedure, if is obtained the preference coefficient $\mu_i (i = \overline{1, m})$ of each sub-objective, then we modify the coefficients by

$$\lambda_i = \frac{w_i \mu_i}{\sum_{i=1}^{m} w_i \mu_i}, i = 1, 2, \cdots, m \quad (4.9)$$

By using weight coefficient of sub-objective, multiple objectives made with one objective by comprehensive method of grey multiple objectives, that is, we obtain such as

$$f(x, \otimes) = \sum_{i=1}^{m} w_i f_i(x, \otimes_1^{(i)}).$$

Then, grey multiple objective programming (4.1) is replaced by

$$\max f(x,\otimes) = \sum_{i=1}^{m} w_i f_i(x,\otimes_1^{(i)})$$

$$s.t. \begin{cases} g_j(x,\otimes_2^{(j)}) \leq 0,\ j=1,2,\cdots,p \\ h_k(x,\otimes_3^{(k)}) = 0,\ k=1,2,\cdots,q \\ x = (x_1, x_2 \cdots x_n) \geq 0 \end{cases} \quad (4.10)$$

The solution of grey one objective programming (4.10) can be obtained by method of Liu Si-feng, Lin Yi (2004).

When $w_i (i = \overline{1,m})$, optimal solution of positioned programming to grey one objective programming (4.10) became to effective solution of positioned programming to multi-objective programming (4.1). The value which optimal point of positioned programming of grey one objective programming (4.10) is placed under each sub-objective function immediately, become to effective value positioned programming corresponds to grey multi-objective programming (4.1).

### 4.2. Algorithm of grey multiple objective programming based on $\theta$-positioned programming method

This method is called by Algorithm 2.

First, we find admissible domain corresponds to $\theta$- weight constrain of given grey multi-objective programming.

Based on calculate process of algorithm 1, we evaluate weight coefficient $w_i (i = \overline{1,m})$ of each objective. If need were, we modify weight coefficients.

Evaluating $\theta$-positioned programming of given grey multi-objective programming (4.1), we obtain ordinary multi-objective programming such as;

$$\max(f_1(x,\otimes_1^{(1)}), f_2(x,\otimes_1^{(2)}), \cdots, f_m(x,\otimes_1^{(m)}))^T$$

$$s.t. \begin{cases} g_j(x,\otimes_2^{(j)}) \leq 0,\ j=1,2,\cdots,p \\ h_k(x,\otimes_3^{(k)}) = 0,\ k=1,2,\cdots,q \\ x \geq 0 \end{cases} \quad (4.11)$$

where $\otimes$ is $\theta$-whitening vector corresponding to grey parameter vector.

We find optimal solution $x^{(i)} = (x_{i1}, x_{i2}, \cdots, x_{in})$ for one objective programming problem such as;

$$\max f_i(x,\otimes_1^{(i)}) \quad (i=1,2,\cdots,m)$$

$$s.t. \begin{cases} g_j(x,\otimes_2^{(j)}) \leq 0,\ j=1,2,\cdots,p \\ h_k(x,\otimes_3^{(k)}) = 0,\ k=1,2,\cdots,q \\ x \geq 0 \end{cases} \quad (4.12)$$

If we denote

$$f_{is} = f_i(x^{(s)}, \otimes_1^{(i)}),\ \underline{f}_i = \min_{1 \leq S \leq m}\{f_{is}\},\ \overline{f}_i = \max_{1 \leq S \leq m}\{f_{is}\} (i=1,2,\cdots,m; s=1,2,\cdots,m),$$

then we obtain relation as follow; $\forall s$, $\underline{f}_i \leq f_{is} \leq f_{ii} = \overline{f}_i$.

If $x^* = (x_1^*, x_2^* \cdots x_n^*)$ is optimal solution of multi objective programming (4.11), then objective value $f_i(x, \otimes_1^{(i)})$ is satisfied relation such as; $f_i(x, \otimes_1^{(i)}) \in [\underline{f}_i, \overline{f}_i](i=1,2,\cdots,m)$.

**Definition 9.** We assume that $w_i (i = \overline{1, m})$ is weight coefficient of each objective in grey multi-objective programming (4.1). We put $\overline{w} = \dfrac{1}{m}$ ( $m$ : number of sub-objective) and denote $w'_i = w_i - \overline{w_i} (i = 1, 2, \cdots, m)$.

If $w'_i \geq 0$, then we obtain

$$\mu(f_i) = \begin{cases} 0 \ ; f_i(x, \otimes_1^{(i)}) < \mu_i + (2w'_i - 1)\nu_i \\ 1 - \dfrac{\mu_i + w'_i \nu_i - f_i(x, \otimes_1^{(i)})}{\nu_i(1 - w'_i)} \ ; \mu_i + (2w'_i - 1)\nu_i \leq f_i(x, \otimes_1^{(i)}) < \mu_i + w'_i \nu_i \\ 1 \ ; f_i(x, \otimes_1^{(i)}) \geq \mu_i + w'_i \nu_i \end{cases} \quad (4.13)$$

If $w'_i < 0$, then we obtain

$$\mu(f_i) = \begin{cases} 0 \ ; f_i(x, \otimes_1^{(i)}) < \mu_i - \nu_i \\ 1 - \dfrac{\mu_i + w'_i \nu_i - f_i(x, \otimes_1^{(i)})}{\nu_i(1 + w'_i)} \ ; \mu_i - \nu_i \leq f_i(x, \otimes_1^{(i)}) < \mu_i + w'_i \nu_i \\ 1 \ ; f_i(x, \otimes_1^{(i)}) \geq \mu_i + w'_i \nu_i \end{cases} \quad (4.14)$$

Then $\mu(f_i)$ is called whitening weight function which sub-objective $f_i(x, \otimes_1^{(i)})$ take optimal value.

Where $\mu_i = \dfrac{1}{2}[\overline{f}_i + \underline{f}_i]$, $\nu_i = \dfrac{1}{2}[\overline{f}_i - \underline{f}_i]$ $(i = 1, 2, \cdots, m)$

**Remark.**

($a$) If there is $i$ which $\overline{f}_i = \underline{f}_i$, then sub-objective $f_i(x, \otimes_1^{(i)})$ can't be considered to find the solution because sub-objective $f_i(x, \otimes_1^{(i)})$ don't give the effect to find optimal solution of multi-objective programming (4.11).

($b$) In above-mentioned composition idea of whitening weight function (for example, formula (4.12), (4.13)) if weight of sub-objective $f_i(x, \otimes_1^{(i)})$ is more great, then value $\mu_i + w'_i \nu_i$ more approach to $\overline{f}_i$ and value range which $f_i(x, \otimes_1^{(i)})$ take is more less relatively.

We assume that $U$ is admissible domain of $\theta$-positioned programming (M-3) and $x \in U$, and define function to minimize over $U$ by

$$F(x) = \mu(f_1) \wedge \mu(f_2) \wedge \cdots \wedge \mu(f_m), x \in U \quad (4.15)$$

Then, easily, we can know that $\forall \lambda \in [0,1], F(x) \geq \lambda \Leftrightarrow \mu(f_i) \geq \lambda (i = 1, 2, \cdots, m)$.

Therefore, the problem to find solution of $\theta$-positioned programming (M-3) immediately, come down to find solution of one objective programming such as;

$$\max G = \lambda$$
$$s.t. \begin{cases} f_i(x, \otimes_1^{(i)}) - (1 - w'_i)\nu_i \lambda \geq \mu_i + (2w'_i - 1)\nu_i, i \in I_1 \\ f_i(x, \otimes_1^{(i)}) - (1 + w'_i)\nu_i \lambda \geq \mu_i - \nu_i, i \in I_2 \\ g_j(x, \otimes_2^{(j)}) \leq 0, j = 1, 2, \cdots, p \\ h_k(x, \otimes_3^{(k)}) = 0, k = 1, 2, \cdots, q \\ x \geq 0 \end{cases} \quad (4.16)$$

Where $I_1 = \{i | 1 \leq i \leq m, w'_i \geq 0\}, I_2 = \{i | 1 \leq i \leq m, w'_i < 0\}$.

Finding solution of problem (4.16) by usual one-objective programming, and getting the value which each objective is placed, then immediately optimal value of each objective is obtained.

## 5. An illustrative example

Calculating Procedure is given such as;

Step 1. 1° Input original data:
  M: the total amount of investment finance,
  i : Number of sort of investment plans,
  $\otimes r_i$ : average profit rate,
  $\otimes q_i$ : risk expense rate when assert $S_i$ is purchased,
  $\otimes p_i$ : transaction cost rate demanded when assert $S_i$ is purchased,
  $\otimes r_0$ : interest rate of Bank,
  $\otimes u_i$ : purchased bound,
  $\otimes \lambda$ : total amount of invest risk determined by investor

Step 2. Evaluation of positioned coefficient of price $\rho$, positioned coefficient of resource restrict $\beta$, positioned coefficient of consumption $\delta$, respectively.

Step 3. Input the grey target $D = [\mu_0, 1]$ :

Step 4. Evaluation of the optimal value $\max \overline{Z}$ of ideal model

Step 5. Evaluation of the optimal value $\max \underline{Z}$ of critical model

Step 6. Evaluation of the positioned programming $LP(\rho, \beta, \delta)$

Step 7. Evaluation of the pleased degree $\mu(\rho, \beta, \delta)$ for positioned programming $LP(\rho, \beta, \delta)$

Step 8. Estimation of the pleased degree $\mu$ whether $\mu \in D$

Step 9. If $\mu \notin D$, then input new $\otimes \lambda$ and go to step 2°

Let us consider grey two objectives programming such as;

$$\max(f_1 = c_{11}(\otimes)x_1 + c_{12}((\otimes)x_2, f_2 = c_{21}(\otimes)x_1 + c_{22}((\otimes)x_2))^T$$

$$s.t. \begin{cases} a_{11}(\otimes)x_1 + a_{12}((\otimes)x_2 \leq b_1(\otimes) \\ a_{21}(\otimes)x_1 + a_{22}((\otimes)x_2 \leq b_2(\otimes) \\ x_1, x_2 \geq 0 \end{cases}$$

Where $c_{11}(\otimes) \in [0, 2], c_{12}(\otimes) \in [1.5, 2.5], c_{21}(\otimes) \in [2, 4], c_{22}(\otimes) \in [-1.5, -0.5]$ ,
$a_{11}(\otimes) \in [2, 4], a_{12}(\otimes) \in [1.5, 2.5], a_{21}(\otimes) \in [-2, 0], a_{22}(\otimes) \in [3, 5], b_1(\otimes) \in [16, 20], b_2(\otimes) \in [7, 9]$.
Let find the solution by algorithm 1.

First, we take mean whitening of constraint conditions, and then take three points $x_1(2,1)$, $x_2(4,2), x_3(5,1)$ in admissible domain. Then, we evaluate each objective value at each point.

Table 1. Placed value of admissible solution of each objective

| objective | $x_1$ | $x_2$ | $x_3$ |
|---|---|---|---|
| $f_1$ | [1.5,6.5] | [3,13] | [1.5,12.5] |
| $f_2$ | [2.5,7.5] | [5,15] | [8.5,19.5] |

Even though we assume that two objective all are effectiveness, we don't lose generality.
Through evaluating, entropy weighting of each objective is obtained such as; $w_1 = 0.6, w_2 = 0.6$.

Transiting the given grey two objectives programming to grey one objective programming, we obtain such as programming;

$$\max f = c_1(\otimes)x_1 + c_2(\otimes)x_2$$

$$s.t. \begin{cases} a_{11}(\otimes)x_1 + a_{12}((\otimes)x_2 \leq b_1(\otimes) \\ a_{21}(\otimes)x_1 + a_{22}((\otimes)x_2 \leq b_2(\otimes) \\ x_1, x_2 \geq 0 \end{cases}$$

Where $c_1(\otimes) \in [0.8, 2.8], c_2(\otimes) \in [0.3, 1.3]$.

By using algorithm of grey one objective programming, optimal solution of mean whitening programming for this programming easily is obtain such as ; $(x_1, x_2) = (6, 0)$.

Optimal value corresponded to it are given by $f_1 = (x_1 + 2x_2)\big|_{(6,0)} = 6$, $f_2 = (3x_1 - x_2)\big|_{(6,0)} = 18$.

Therefore, effective solution of mean positioned programming for the given grey two objective programming is given by $(x_1, x_2) = (6, 0)$. The efficient value of each objective function is $f_1 = 6$, $f_2 = 18$.

By similar method, we can obtain the effective solution of other positioned programming for the given grey two-objective.

Let find the solution by algorithm 2.

Let's find only the solution of mean whitening programming (for case $\theta = 0.5$) for given two objective programming.

By evaluating the analogous method with algorithm 1, we obtain each objective weight $w_1 = 0.6$, $w_2 = 0.4$.

After taking the mean whitening for grey parameters, we have single objective programming such as;

$$\max G = \lambda$$

$$\begin{cases} x_1 + 2x_2 - 1.8\lambda \geq 6.4 \\ 3x_1 - x_2 - 4\lambda \geq 9 \\ 3x_1 + 2x_2 \leq 18 \\ -x_1 + 4x_2 \leq 8 \\ x_1, x_2, \lambda \geq 0 \end{cases}$$

Solving this programming, we have optimal solution; $x = (x_1, x_2, x_3) = [\frac{34.2}{7}, \frac{11.6}{7}, 1]$. That is, $x_1 = \frac{34.2}{7}, x_2 = \frac{11.6}{7}$ is optimal solution of mean whitening programming of given grey two objectives programming under the degree of confidence $\lambda = 1$. Optimal value corresponded to each objective are given by $f_1(\frac{34.2}{7}, \frac{11.6}{7}) = 8.2, f_2(\frac{34.2}{7}, \frac{11.6}{7}) = 13$.

In this paper, we have defined grey multi objective programming under the general meaning and given the method to determine the weight of objective, and established two kind of algorithm to fond grey multi objective programming. Algorithm 1 has universal meaning. Algorithm 2 has accepted whitening weight function to take for sub-objective optimal value by using weight of sub-objective. Furthermore, we have diverted $\theta$-positioned programming of grey multi objective programming to general single objective programming. These algorithms all regard as of great importance uncertainty (greyness) at grey multi objective programming and simple and easy the calculating process. The calculating examples of paper also show ability and effectiveness of algorithms.